\documentclass[preprint,aps,12pt]{revtex4}
\usepackage{graphicx}

\newif\ifpdf\ifx\pdfoutput\undefined\pdffalse\else\pdfoutput=1\pdftrue\fi

\begin{document}

\title{Sharp vorticity gradients in two-dimensional hydrodynamic turbulence }
\author{E. A. Kuznetsov}
\affiliation{L.D.~Landau Institute for Theoretical Physics,
2 Kosygin str., 119334 Moscow, Russia}
\author{V. Naulin}
\author{A.~H. Nielsen}
\author{J.~Juul Rasmussen}
\affiliation{Optics and Plasma Research Department, OPL-128, Ris{\o}
National Laboratory, P.O.Box 49, DK-4000 Roskilde,  Denmark }

\date{2004-07-16}

\begin{abstract}
The appearance of sharp vorticity gradients in two-dimensional
hydrodynamic turbulence and their influence on the turbulent spectra
is considered. We have developed the analog of the vortex line
representation as a transformation to the curvilinear system of
coordinates moving together with the di-vorticity lines.
Compressibility of this  mapping can be considered as the main
reason for the formation of the vorticity discontinuities at high
Reynolds numbers. For two-dimensional turbulence in the case of
strong anisotropy the vorticity discontinuities can generate spectra
with the fall-off at large $k$ proportional to  $k^{-3}$ resembling the Kraichnan
spectrum for the enstrophy cascade. For turbulence with weak
anisotropy the $k$ dependence of the spectrum due to discontinuities
coincides with that of the Saffman spectrum: $k^{-4}$. We have
compared the analytical predictions with direct numerical solutions
of the two-dimensional Euler equation for decaying turbulence. We
observe that the di-vorticity is reaching very high values and is
distributed locally in space along piecewise straight lines. Thus,
indicating strong anisotropy and accordingly we found a spectrum
close to the $k^{-3}$-spectrum.
\end{abstract}




\maketitle

\section{Introduction}

This paper is concerned with investigations of two-dimensional (2D)
hydrodynamical turbulent flows. In particular, we study the
formation and dynamics of very sharp vorticity gradients and their
influence on the energy spectrum in the enstrophy cascade regime. We
may consider two kind of turbulent spectra. The first one was
suggested by Kraichnan in 1967 \cite{kraichnan}, it corresponds to
the enstrophy cascade directed to the small-scale region where
viscous dissipation becomes essential. The Kraichnan spectrum follows,
up to the logarithmic factor (see \cite{kraichnan1}), a power law
for the scales intermediate between source and sink (the inertial
interval): $E(k)\sim\eta^{2/3} k^{-3}$ where $\eta$ is the
enstrophy dissipation rate. Recall that 2D turbulence additionally
is characterized by an inverse energy cascade for large-scales
leading to $E(k)\sim k^{-5/3}$ (see e.g. \cite{kraichnan}). However,
in the present paper we will only be concerned with the small-scale
region of the spectrum. The second spectrum suggested by Saffman in
1971 \cite{saffman} yields another power dependence: $E(k)\sim
k^{-4}$. According to Saffman, in decaying 2D turbulence vorticity
discontinuities (in absence of viscosity) will form because fluid
elements with different values of vorticity will be driven close
together by the flow. Due to vorticity conservation the appearance
of discontinuities will provide the conservation of all other
invariants involving vorticity, $\int\omega^ndS$, $n=3,4,...$.
Accounting for a finite viscosity Saffman considers the
"discontinuities" to have a small width $\delta$, which results from
the balance between inertial and viscous forces. At high-Reynolds
number this size is assumed to be very small in comparison with the
length along the discontinuities, $L$, which may be assumed to be of
the same order as the characteristic energy-containing length-scale.
Under the  assumption of isotropy and a dilute distribution of
discontinuities Saffman suggested that the energy spectrum at large
$k$ could be constructed as a superposition of the spectra from the
individual discontinuities resulting in: $E(k)\sim k^{-4}$ .

From the first sight, the spectra obtained by Kraichnan and  Saffman
look like two different answers, but indeed, as we show in this
paper,  it is possible to establish some connection between them.
This may be seen from the Fourier transform of a step function. Let
us assume that the vorticity has a jump $\Gamma=\Gamma(x)$ along the
line $y=0$ and at first neglect effects connected with bending of
the line. Then we can write:
$$
\frac{\partial\omega}{\partial y}=\Gamma\delta (y).
$$
Hence it is immediately seen that the Fourier transform will have a
power-law fall-off at large $k$, i.e., inversely proportional to
$k_y$ multiplied by some function of $k_x$ due to dependence of
$\Gamma$ on $x$.  If we neglect the dependence on $k_x$ replacing it
by some constant, then we immediately obtain an energy spectrum with
a power dependence similar to the Kraichnan spectrum: $E(k)\sim
k^{-3}$.
 This is an important conjecture demonstrating that a spectrum similar to the Kraichnan
 spectrum, which is often observed in high resolution numerical
 simulations, may be related to discontinuities of vorticity which can be considered as possible candidates
for singularities in  ideal fluids in two dimensions. However, for
viscous fluids, i.e., in the
 framework of the Navier-Stokes equation,
such singularities are impossible: the initial smoothness of the
solution will remain as proven first by Olga Ladyzhenskaya many years
ago~\cite{lad}. 
Within the 2D Euler equations for incompressible fluids
the  vorticity is a Lagrangian invariant and can never be singular,
but its gradient might, in principle, become infinite in a finite
time. Up to now this is an open question.

It is necessary to mention some examples of 2D flows presented by
Yudovich \cite{yudovich}, where the appearance of weaker
singularities (vorticity is allowed to be discontinuous but bounded)
are possible, however, they are formed in infinite time.  Another
approach based on the numerical analysis of the complex
singularities for the inviscid flow with two-mode initial conditions
showed that the width of its analyticity strip follows a
$\mbox{ln}(1/t)$ law at short times \cite{FMB,MBF}. 
Additionally, many numerical experiments for 2D turbulence (see,
\cite{mcwilliams}- \cite{benzi}) show that with a good accuracy the
Saffman spectrum is formed at the initial stage, before the
excitation of the long-scale coherent vortices. The high-resolution
numerical simulation performed by Legras et al.~ \cite{legras}
demonstrated the power dependence $k^{-3.5}$. Analytical
calculations presented by Gilbert \cite{gilbert} using arguments based
on the existence of spiral
structures  give a power dependence with exponent between
$-3$ and $-4$ (see also \cite{moffatt} and \cite{vasilicos}). In
particular, we would like to point to the very interesting paper by
Ohkitani \cite{okhitani}, where  by means of the Weiss decomposition
\cite{weiss} it was shown that the so-called {\it h}-regions ({\it
h} - hyperbolic in the sense of Ref. \cite{weiss}, i.e., regions
where straining is dominating over vorticity) give the spectrum
$k^{-3}$, i.e., coinciding up to a logarithmic factor with the
Kraichnan spectrum, the contribution from the {\it e}-regions ({\it
e} - elliptic, i.e., vorticity dominated regions) yield 
the Saffman spectrum $\sim k^{-4}$. Note, that similar
ideas based on the wavelet analysis were developed in \cite{DBPD} to
separate 2D turbulent flow into regions having different dynamical
behaviors. The appearance of a power type spectrum in the short-wave
region has been connected with different physical mechanisms like
vortex merging \cite{NHRB}, \cite{RNN} and vortex stripping
\cite{dritschel}, \cite{legrasdritschel} which give a certain
confirmation of the original idea of Saffman \cite{saffman}.

In this paper we present some qualitative physical arguments in
favor of the formation of vorticity discontinuities in the 2D Euler
equations for smooth initial conditions.  The main idea in the
description of 2D flows is connected with using the vorticity as
Lagrangian invariants. Kuznetsov and Ruban \cite{KR} (see also
\cite{kuz}) developed a new kind of description for
three-dimensional vortical flows - the so-called vortex line
representation (VLR). This representation is based on the mixed
Lagrangian-Eulerian  description and connected with movable vortex
lines. The VLR, which is a mapping to a curvilinear system of
coordinates, turns out to be compressible, this is considered to be
the main reason for breaking in hydrodynamics. Here we demonstrate
how this approach can be  modified for 2D Euler hydrodynamics.
The main observation is that for 2D flows the curl of the vorticity,
sometimes referred to as the di-vorticity
\cite{weiss,kida,okhitani}, represents a frozen-in field, i.e., it
satisfies  the same equation as, e.g., the equation for the magnetic
field in ideal 2D magneto-hydrodynamics (MHD). Therefore the
generalization to the 2D Euler equations becomes straightforward. In
the local case, as it was demonstrated for 2D MHD in Ref.
\cite{KPS}, the vorticity plays the role of a Lagrangian coordinate
and the other variable coincides with the Cartesian coordinate, say,
$x$. In terms of these variables \cite{KPS} the 2D Euler equations
transform into equations of motion for a layered fluid, similar to
stratified fluid, where each layer is labeled by its vorticity
$\omega$. In terms of the new variables the "new" hydrodynamics
becomes compressible. The derivative $y_{\omega}$ plays the role of
density of each layer, as a function of time and coordinate  $x$.
This characteristics is proportional to the width between two
neighboring layers with closed vorticity contours.

Another aim of this paper is to revisit the energy spectra for 2D
turbulence with emphasis on the angle distribution, following the
arguments of Saffman connected with vorticity discontinuities. Using
the stationary phase method we demonstrate that the contribution
from one  discontinuity is very anisotropic: it has a sharp angular
peak along the direction perpendicular to the discontinuity. In the
peak the energy spectrum falls-off like $k^{-3}$ at large $k$. After
average over angles in the case of isotropic turbulence the spectrum
coincides with the Saffman spectrum \cite{saffman}.

In order to check whether the spectrum at large $k$ is defined by
vorticity discontinuities we have performed numerical experiments on
decaying turbulence based on a direct numerical solution of the 2D
Euler equations. In the turbulent state when the formation of power
tails is observed we examine the structure of the di-vorticity. We
found that the di-vorticity is distributed very sharply in space
concentrated on a random net of lines. In our opinion, these results
can be interpreted in favor of the Saffman mechanism for the
formation of 2D turbulent spectra due to discontinuities.

\section{Two-dimensional analog of the VLR}

Consider  a 2D ideal fluid, described by the Euler equation for the
vorticity  $\omega(x,y,t)$,
\begin{equation}
\label{omega-t} \frac{\partial \omega}{\partial t}+({\bf v}\cdot
\nabla)\omega=0,\,\,\, \mbox{div} \, {\bf v}=0,
\end{equation}
where the velocity field ${\bf v}$ defines the vorticity:
$$
\omega= \nabla \times {\bf v} = \frac{\partial v_y}{\partial x}-
\frac{\partial v_x}{\partial y}.
$$
Equation (\ref{omega-t}) shows that the vorticity is a Lagrangian
invariant advected by the fluid, i.e.,
$$
\omega(x,y,t)=\mbox{inv}
$$
along a fluid particle trajectory defined as solution of the system
of ordinary differential equations (ODE's),
\begin{equation}
\label{map}
\frac{d\bf r}{dt}={\bf v(r,}t),\,\, {\bf r}|_{t=0}={\bf a}.
\end{equation}

Let us introduce the divergence-free vector field
${\bf B}$ with the components
$$
B_x=\frac{\partial \omega}{\partial y},\,\, B_y=-\frac{\partial
\omega}{\partial x},
$$
i.e., ${\bf B} = \mbox{curl} \, \omega \hat{z}$. It is easily to see
that this vector is tangent to the line $\omega(x,y)=\mbox{const}$
because the vorticity gradient $\nabla \omega=(\partial_x\omega,
\partial_y\omega)$ is normal to this line. The equation of motion
for the vector ${\bf B}$ can be obtained from Eq. (\ref{omega-t})
after differentiating with respect to coordinates:
\begin{equation}
\label{divort} \frac{\partial {\bf B}}{\partial t} = \mbox{curl}\,
[{\bf v}\times{\bf B}].
\end{equation}
Thus, the vector ${\bf B}$ constitutes a frozen-in quantity.
Sometimes, it is called as the di-vorticity (see \cite{kida}). By
introducing the substantial (material) derivative,
$d/dt=\partial/\partial t+ ({\bf v}\cdot \nabla)$, Eq.
(\ref{divort1}) can be rewritten as
\begin{equation}
\label{divort1}
\frac{d\bf B}{dt}=({\bf B}\cdot \nabla){\bf v}.
\end{equation}
Hence, we observe that $|{\bf B}|$ will locally increase due to
stretching of the di-vorticity lines, i.e., when
\begin{equation}
\label{stretching}
\frac 12\frac{d\bf B^2}{dt}=({\bf B}\cdot {\hat S}{\bf B}) >0,
\end{equation}
where
$$
\hat S_{ik}=\frac 12 \left( \frac {\partial v_k}{\partial x_i}+
\frac {\partial v_i}{\partial x_k}\right)
$$
is the stress tensor. Increasing (or decreasing) the di-vorticity
field, based on the equation (\ref{stretching}), is not sufficient
to clarify  the physical mechanism  for its growth. As is seen from
Eq. (\ref{divort}) only one velocity component, ${\bf v}_n$, normal
to the vector ${\bf B}$  changes  the field ${\bf B}$. In this case
the tangential component ${\bf v}_{\tau}$ (parallel to ${\bf B}$)
plays a passive role providing the incompressibility condition:
$\mbox{div}\,{\bf v}_{\tau}+ \mbox{div}\,{\bf v}_n=0$.
 This observation is the key point
for introducing the vortex line representation (VLR) for the
three-dimensional Euler equations (see, e.g. \cite{kuz}). To
construct the analog of VLR for the 2D Euler equations  we consider
new Lagrangian trajectories, given by the ${\bf v}_n$,
\begin{equation}
\label{map2}
\frac{d \bf r}{dt}={\bf v}_n({\bf r},t), \,\,\,{\bf r}|_{t=0}={\bf a}.
\end{equation}
The solution of these ODE's defines a new mapping
\begin{equation}
\label{map1}
{\bf r}={\bf r}({\bf a},t)
\end{equation}
which is different from that given by Eq. (\ref{map}). In terms of
this mapping the di-vorticity equation (\ref{divort}) can be
integrated (for details see, e.g. \cite{KPS}):
\begin{equation}
\label{Bfield}
{\bf B(r},t)=\frac{({\bf B_0(a)}\cdot\nabla_a){\bf r}({\bf a},t)}{J},
\end{equation}
where ${\bf B_0(a)}$ is the initial di-vorticity,  $J$ is the
Jacobian of the mapping (\ref{map1}):
$$
J=\frac{\partial(x,y)}{\partial(a_x,a_y)}.
$$
According to the definition of this mapping its Jacobian is not
fixed, it may change in time and space. In other words, the mapping
${\bf r}={\bf r}({\bf a},t)$, as a change of variables, represents a
compressible mapping. This means that the di-vorticity lines can be
compressed. In this approach the velocity of motion of di-vorticity
lines is nothing else than the normal velocity ${\bf v}_n$.

It is interesting to note that this approach in slightly different
form what was suggested in \cite{KPS}. In this paper the basis of
the approach is the mixed Lagrangian-Eulerian description when all
desired quantities are considered as functions of vorticity $\omega$
(or any other Lagrangian invariant) and a Cartesian coordinate $x$.

The VLR given by (\ref{map2}), (\ref{map1}), (\ref{Bfield}) with the local 
change of variables ${\bf r}={\bf r}({\bf a},t)$, does not work at singular  
points where the ${\bf B}$-field vanishes,
\begin{equation} 
\label{zero} 
{\bf B}({\bf r}(t),t)=0.  
\end{equation}  
and where, respectively, the normal 
velocity is not defined. For vorticity $\omega$ these points are nothing more
than maximal, minimal or saddle points. 
It is easy to see that the null points are advected by the fluid, but the
velocity  ${\bf v}$ at these points is defined through the ${\bf B}$-field 
by inverting the  
Laplacian operator: ${\bf v}=-\Delta^{-1}{\bf B}$. 
The null-points for the normal vector  
field ${\bf n}({\bf r})$ represent topological singularities. 
Topological constraints as additional conditions 
to the system (\ref{map2}), (\ref{map1}), (\ref{Bfield}), 
are written as integrals of the vector field  
${\bf n}({\bf r})$ 
along a loop  enclosing the null-points:  
\begin{equation} 
\label{con2} 
\oint (\nabla \varphi\cdot d{\bf r}) =2\pi m,  
\end{equation} 
where $\varphi$ is the angle between the vector 
${\bf n}$ and the $x$-axis and  
$m$, being a topological charge,  is an integer   
equal to the total number of turns of the vector ${\bf n}$  
while passing around the closed contour with the null-point inside it
(see also \cite{KPS}). For instance, for $X$-points or $O$-points, $m=\pm 1$ .  

%


As well known from our knowledge in  gas-dynamics compressibility of
the mapping is a main cause for steepening and ultimately breaking,
resulting in the formation of sharp gradients for the velocity and
density of the gases. This happens in finite time and in the general
situation the singularity first appears in one separate point, i.e.,
it may be related to collapse. In gas-dynamics this process is
completely characterized by the mapping determined by the transition
from the Eulerian to the Lagrangian description. Vanishing of the
Jacobian corresponds to the emergence of a singularity. For
three-dimensional incompressible Euler equations compressibility of
the VLR is a possible reason for appearance of infinite vorticity in
one separate point that results in breaking of vortex lines. The
first study of vortex-line breaking for three-dimensional integrable
hydrodynamics with the Hamiltonian $\int|{\bf \omega}| d{\bf r}$ was
performed by Kuznetsov and Ruban \cite{collapse}. Recent numerical
experiments \cite{ZKP}, \cite{KPZ} have confirmed the possibility of
this type of scenario.

The Jacobian in dominator of the expressions (\ref{Bfield}) can
become zero, which will result in infinite value of the
di-vorticity. We do not see any restrictions by which this process
can be forbidden. In 2D hydrodynamics, however, compressibility of
the mapping guarantees only compression of di-vorticity lines
corresponding to the
 formation of sharp gradients for vorticity. Probably,
the breaking process in 2D happens in infinite time (see, for
instance, \cite{yudovich}). The most important point for us is that
the tendency indeed does exist and it is possible to imagine that
this process may be accelerated in the presence of external forces
driving the turbulence.

\section{2D spectrum}

In the previous section we gave some arguments in favor of formation
of sharp gradients of the vorticity in 2D Euler flows. Everywhere
below we will suppose that this process is possible and consider how
it can effect the form of turbulent spectrum. For 2D turbulence, in
the presence of finite viscosity and external forces, we will assume
that the sharp vorticity gradients have a finite value inversely
proportional to the characteristic width of discontinuity $\delta$,
which is defined from the balance between inertial and viscous
terms. At high Reynolds number the width $\delta$ will be much less
than the characteristic (energy-containing) scale $L$. In the
turbulent state such discontinuities are naturally assumed to form a
set of vorticity gaps with random positions of their centers, random
forms and random distributions over angles. Our aim is to calculate
the contribution to the spectrum from such discontinuities. We will
be interested in the region of wave numbers $k$ lying between
$L^{-1}$ and the inverse width $\delta^{-1}$:
$$
L^{-1}\ll k\ll \delta^{-1}.
$$
To simplify the problem all gaps are supposed to be concentrated on
pieces of straight lines (finite intervals) with vorticity gaps
vanishing at the endpoints of the intervals. As we will see later
this simplification is not so essential. The answer, which we will
get, will also account for bends of the discontinuity lines.

To find spectrum we need to calculate the Fourier transform from of
pair correlation function:
$$
F({\bf r})=\langle\omega({\bf x})\omega({\bf x}+{\bf r})\rangle,
$$
where angle brackets means average over the ensemble of
discontinuities. Hence the energy density spectrum $\epsilon ({\bf
k})$ is given by the standard formula:
$$
\epsilon({\bf k})=\frac{F_k}{2k^2} =\frac{\overline{|\omega_k|^2}}{8\pi^2Sk^2}
$$
where $\omega_k$ is the Fourier transform of the vorticity
$\omega({\bf r})$,
$$
\omega_k= \int \omega({\bf r})e^{-i({\bf k r})}d{\bf r},\,\,\,
\omega({\bf r})=\frac{1}{(2\pi)^2}\int \omega_k e^{i({\bf k r})}d{\bf k},
$$
the over-bar denotes average with respect to random variables, and
$S$ is the average area, which is assumed to be sufficiently large.

Consider first one discontinuity with the center at ${\bf
r_0}=(x_0,y_0)$ oriented along the $y$-axis. Then for the
y-derivative of $\omega$ we have,
\begin{equation}
\label{gap}
\frac{\partial\omega}{\partial y}=\Gamma(x)\delta(y-y_0)+\,\mbox{regular terms}.
\end{equation}
Here $\Gamma(x)$ is a continuous function of $x$ inside the interval $[x_1,x_2]$ vanishing
at the endpoints $x=x_{1,2}$ and equal zero outside the interval.

Hence, the Fourier transform from the singular part of $\omega$ is
given by the integral:
$$
\omega_k=-\frac{i}{k_y} e^{-ik_yy_0} \int_{x_1}^{x_2}\Gamma (x) e^{-ik_xx} dx
$$
where ${\bf k}=(k_x,k_y)$. This is the contribution from one
discontinuity. If we assume that the discontinuities are not very
densely distributed, they may be considered "independent" and the
spectrum for the whole ensemble of discontinuities may be obtained
by a superposition of the spectra from the individual
discontinuities, i.e., from the summation with respect to all
discontinuities which results in
\begin{equation}
\label{sum} \omega_k=-\sum _{\alpha}\frac{i}{({\bf k\cdot
n_{\alpha}})}e^{-i({\bf k\cdot n_{\alpha}})y_{0\alpha}}
\int_{x_{1\alpha}}^{x_{2\alpha}} \Gamma_{\alpha}(x) e^{-i({\bf k
\cdot \tau_{\alpha}})x} dx.
\end{equation}
Here we have introduced two unit vectors: normal ${\bf n_{\alpha}}$
and tangent ${\bf \tau_{\alpha}}$ (${\bf n_{\alpha}}^2={\bf
\tau_{\alpha}}^2=1,\,\, ({\bf n_{\alpha} \tau_{\alpha}})=0$)
characterizing the
 orientation  of the $\alpha$-th discontinuity. The coordinates $x_{1\alpha},x_{2\alpha}, y_{0\alpha}$
together with the two unit vectors define completely the positions
of the discontinuities.

To find an enstrophy spectrum one needs to perform average of
$|\omega_k|^2$ with respect to all random variables. Assuming the
coordinates of the discontinuities to be randomly distributed
uniformly in space, the first average gives:
\begin{equation}
\label{dist}
\overline{|\omega_k|^2}= N\langle \frac{1}{({\bf k\cdot n})^2}
\left|\int_{x_{1}}^{x_{2}}
\Gamma(x) e^{-i({\bf k\cdot\tau})x} dx\right|^2\rangle.
\end{equation}
Here $N$ is the number of discontinuities  in area $S$,
angle brackets means the average with respect to both $x_1,x_2$ and 
angle distribution.

Since we are interested in short-wave asymptotics of the spectrum,
$kL\gg 1$, the integrand  in (\ref{dist}) in this case represents a
rapidly varying function of x. Therefore
 the integral  in (\ref{dist}) can be estimated by means of the method 
 of stationary phase.
 This method can be applied for all angles except for a narrow cone of 
 angles, $\theta_k$ ($\theta_k$
 is the angle between the vectors ${\bf n}$ and ${\bf k}$)
 where $kL\theta_k\leq 1$.
 In this region the integral can be considered as constant which results in
 the following form for the energy distribution (before angle averaging!):
 \begin{equation}
\label{small}
\epsilon_1({\bf k})\approx \frac{n}{8\pi^2k^4}\langle
\left(\overline{\Gamma}l\right)^2\rangle,\,\,\, \theta_k\leq \theta_0\equiv (kL)^{-1},
 \end{equation}
 where $n$ is the density of discontinuities ($=N/S$) and
$$
\overline{\Gamma}l= \int_{x_{1}}^{x_{2}}
\Gamma(x) dx, \,\,\, l=x_2-x_1, \,\, \langle l\rangle=L.
$$
For angles $\theta_k$ lying far from $\theta_0\equiv (kL)^{-1}$ the integral in  (\ref{dist})
can be estimated by means  of the method of stationary phase. However,
the leading order, proportional to $({\bf k\tau})^{-1}$, gives zero input because
$\Gamma(x_{1,2})=0$. Therefore one needs to keep the next order approximation that gives:
\begin{equation}
\label{large}
\epsilon_2({\bf k})\approx \frac{N}{4\pi^2k^2}\frac{\langle(\Gamma')^2\rangle}
{({\bf k\cdot n})^2({\bf k\cdot \tau})^4}
,\,\,\, \theta_k\gg (kL)^{-1},
 \end{equation}
where $\Gamma'$  is the derivative of $\Gamma$ taken at the
endpoints $x_{i}$. This formula demonstrates singular behavior for
$\epsilon({\bf k})$ at angles $\theta_k$ close to $0$ and $\pi/2$ (as well as,
to $\pi$ and $-\pi/2$).
At small angles  $\theta_k\leq (kL)^{-1}$ this expression has to be
matched with (\ref{small}). For the angle range close to $\pi/2$ the
integral in (\ref{large}) should be cut-off due to the bending of
the line of discontinuity. This  factor switches on at angles
$|\theta_k-\pi/2|\sim (ka)^{-1}$ where $a$ is a characteristic
bending length of discontinuity (roughly of the order of $L$). Thus,
the energy density distribution $\epsilon({\bf k})$ has a very
narrow angle maximum at $\theta_k$ near zero with decay at large
wave numbers as $\sim k^{-4}$, this results in the energy spectrum
$E(k)\sim k^{-3}$, which, up to the logarithmic factor, corresponds
to the Kraichnan spectrum . For all other angles $\epsilon(k)$
decays  proportionally to $k^{-6}$ at large $k$.

We would like to stress once more that the formulas (\ref{small})
and (\ref{large})
 are the results of non-complete average, i.e, the average with respect to
coordinates $x_{1\alpha},x_{2\alpha}, y_{0\alpha}$. In order to get
the final answer for the energy spectrum it is necessary to average
with respect to angles.

Let us assume first that the angle distribution is isotropic. Then,
integrating over angles it is easily seen that from the first region
(\ref{small}) we have the following contribution:
\begin{equation}
\label{smallint}
E_1(k)=2k\int_{-\theta_0} ^{\theta_0} \epsilon_1(k)d \theta_k
\approx\frac{n}{2\pi^2k^4L}\langle
\left(\overline{\Gamma}l\right)^2\rangle.
 \end{equation}
Here the factor 2 appears because of two equal contributions from
two regions near $\theta_k=0$ and $\theta_k=\pi$. Averaging
(\ref{large}) over angles gives divergence at $\theta\to 0$ and
$\theta\to \pi/2$. The main contribution to the energy spectrum
comes from the cut-off at small angles $\sim (kL)^{-1}$:
\begin{equation}
\label{largeint}
E_2(k)\approx\frac{nL^3}{3\pi^2k^4}
\langle(\Gamma')^2\rangle .
\end{equation}
Thus, both regions of angles give contributions of the same order of
magnitude. The complete answer for the energy spectrum for isotropic
turbulence (i.e., isotropic distribution of discontinuities) is given
by the sum of (\ref{smallint}) and (\ref{largeint}):
\begin{equation}
\label{sumenergy} E(k)\approx\frac{n}{2\pi^2k^4L}\left[\langle
\left(\overline{\Gamma}l\right)^2\rangle
+\frac{2L^4}{3}\langle(\Gamma')^2\rangle \right],
\end{equation}
which coincides with the spectrum obtained by Saffman
\cite{saffman}.

In order to find the spectrum in the  anisotropic situation one
needs to average expressions (\ref{small}), (\ref{large}) with the
corresponding distribution function. In numerical experiments
anisotropy can be conditioned by box boundaries as well as by
anisotropy of the pumping of turbulence. In the case when such
ordering is strong enough the spectrum may get some peculiarities
originating from non-averaged spectra given by  (\ref{small}),
(\ref{large}). If the width of the angle distribution function
$\Delta\theta$ will be narrower than $\theta_0=(kL)^{-1}$, then  in
the angle cone $\theta_k <\Delta\theta$ the energy spectrum
$E(k,\theta)$ will have the fall-off $\sim k^{-3}$, i.e., the same
power dependence as for the Kraichnan spectrum. Note, however, that
this asymptotics is only intermediate because $\theta_0=(kL)^{-1}$
decreases with increasing $k$. Therefore starting from
$k=k^{*}=(L\theta_0)^{-1}$, the average over angles becomes
sensitive relative to the singularities of (\ref{large}) that
results in the spectrum
 decreasing proportional to the Saffman fall-off.
If the influence of anisotropy is not so essential then we should
expect the spectrum close to the Saffman one, of course, in the case
when the main contribution to the spectrum at large $k$ is connected
with discontinuities. The most interesting observation following
from the analytical results of this Section is that in the very
anisotropic case with strong ordering of discontinuities the sharp
angular maximum of the spectrum has the power fall-off coinciding
with that for the Kraichnan spectrum. While in the isotropic case
our answer coincides with the Saffman answer. In the next Section we
present the results of numerical simulation of decaying 2D
turbulence at high-Reynolds numbers. In particular, the appearance
of the power law tails in the energy spectrum at large $k$ can be
explained rather by discontinuities, than by a cascading process
with constant enstrophy.

\section{Numerical investigations}
To support the arguments of the previous sections and reveal the
direct connection between the formation of the sharp vorticity
gradients and the tail of the energy spectrum we have performed a
numerical study of the evolution of decaying 2D turbulence. The 2D
Euler equations (\ref{omega-t}) in the vorticity-streamfunction
formulation are integrated numerically on a double periodic domain
by employing a high resolution fully de-aliased spectral scheme:
\begin{equation}
 \frac{\partial \omega}{\partial t} + \left\{\omega , \psi \right\} = \mu_{2n} \nabla^{2n} \omega,
 \label{eq:hpEuler}
\end{equation}
where $\psi$ is the streamfunction related to the vorticity by the
Poisson equation:  $\omega = - \nabla^2 \psi$, the velocity is given
as ${\bf v} = (v_x , \, v_y ) = \nabla \psi \times \hat{z}$ and the
bracket 
$$
\{ \omega, \psi \} \equiv   {\bf v  } \cdot \nabla \omega =
\frac{\partial \omega}{\partial x} \frac{\partial \psi}{\partial y}
- \frac{\partial \psi}{\partial x}\frac{\partial \omega}{\partial
y}.
$$ 
In solving (\ref{eq:hpEuler}) we have included a hyperviscosity
term on the right hand side of the equation to keep the integration
scheme stable (typically we have used $n = 3$ and  $\mu_{6} =
10^{-20}$). This term was observed to decrease the energy by less
than 0.002\% and the enstrophy by less than 20\%. We verified that
the global features of our results were not dependent on the type of
viscosity (alternatively we used kinematic viscosity). In the present
context we apply the hyperviscosity to allow an as wide a dynamical
range as possible with the given resolution. The domain size is
taken to be unity and the resolution is $2048 \times 2048$ modes. For the
time integration we employ a third order stiffly-stable scheme. We
have chosen the time scale to correspond to $\omega_{0}^{-1} $,
where $\omega_0 $ corresponds to the maximum vorticity.

As initial condition we have placed a number of positive and
negative vortices randomly on the domain, ensuring that the total
circulation is zero. Vortices of various
shapes/profiles from vortex patches (Rankine vortices) to Gaussian
vortices form the initial condition. In the simulation run described
here,  we have thus used 10
vortices of each sign with Gaussian profiles:
\begin{equation}
 \omega (r, \theta) =  \pm \omega_0
 \exp{(-r^2 / R_{0}^{2})} , \label{eq:initial}
\end{equation}
where $\omega_0 = 1$ for all vortices, while their radii $R_0 $ are
varying in the range $ 0.02 < R_0 < 0.075 $. In Fig.~1a we show the
initial vorticity field and Fig.~1b the vorticity field at $T
= 95 $, which corresponds to around 8 vortex internal  turnover
times ($T_v \equiv 4\pi /\omega_0 $). The vorticity field has the
typical structure for 2D turbulence; it consist of large scale
structures (vortices) with concentrated vorticity and strongly
filamented structures between the vortices. At this time there is
still strong dynamics in the flow evolution, with shearing and
straining due to mutual interactions of nearby structures.
Corresponding to the vorticity field we show the instantaneous
one-dimensional energy spectrum $E(k)$ (total energy: $E =
\int_{o}^{\infty} E(k) dk$) in Fig. \ref{Fig:spectra}. The spectrum
$E(k)$ for $T=0$ in Fig. \ref{Fig:spectra} shows the spectrum of
superimposed Gaussian vortices. The spectrum is expanding to the
high k-values and at $T = 95  $ a $k^{-\alpha}$ spectrum is
developed at high wavenumbers for the present case $\alpha \sim 3$,
as clearly demonstrated in Fig.~\ref{Fig:spectra}, which show the
compensated spectrum $k^3 E(K) $ being constant over almost 2
decades in $k$.

To investigate the details of the dynamics and how the $k^{-\alpha}$
spectrum is generated, we plot in Fig.~\ref{Fig:divorticity} the
di-vorticity field ${\bf B }$ defined in Sec.~II, showing the length
$|{\bf B }|$, which is equal to $|\nabla \omega |$.
It is clear from the figure that very sharp vorticity gradients
appears. These are localized in stripes that are mostly along
straight lines. The stripes are mainly formed outside the dominating
vortex structures, and their formation can be explained by the analysis
discussed in Sec. II. Furthermore it is evident that the
concentration of the stripes are relatively low. Following the time
evolution of maximum value of $|{\bf B }|$, $B_{max}$, we observe a
very rapid growth to values more than 100 times the initial value.
$B_{max}$ oscillates in time with a typical period related to the
vortex turnover time, $T_v $. The highest maximum reached during
this simulation approach 1000, which with a maximum value of the
vorticity $\omega_0 = 1$ corresponds to the width of the filaments
$\delta < 0.001 $, it is evident that the growth of $|{\bf B }|$ is
arrested by the hyperviscosity, and indeed $B_{max}$ scales with
$\mu_{6}^{-1/6} $.
We compare the structure of the di-vorticity  field with the
high pass filtered vorticity field shown in Fig. 4. The very similar
structure of the high pass filtered vorticity field and the
vorticity gradient field strongly suggests that the vorticity
gradients are responsible for the large-$k$ part of the spectrum,
i.e., the $k^{-\alpha}$ part.

To further discuss the dynamics we show the Weiss field
\cite{weiss} in Fig.~5, defined by $W = \frac{1}{4}(s^2 - \omega^2)$
where  
$$
s = [(\frac{\partial v_x }{\partial x} - \frac{\partial
v_y}{\partial y})^2 + (\frac{\partial v_y }{\partial x} +
\frac{\partial v_x}{\partial y})^2 ]^{1/2}
$$
is the rate of deformation. Comparing Figs. 3 and 5 we observe that 
the vorticity
gradient stripes are aligned with the contours of $W$ in the strain
dominated regions, $W > 0$, mainly at the edge of the vortex
structures and in between the structure. A careful inspection,
however, reveals that in the stripes $W < 0 $, i.e., vorticity
dominates. This is in line with the original arguments of Weiss (see
also \cite{okhitani,CEEWX}) that vorticity gradients will tend to
concentrate in the strain dominated regions. In particular in the
work of Chen {\it et al.}~\cite{CEEWX} it was demonstrated that the
dynamics leading to the enstrophy cascade in driven 2D turbulence is
most significant in strain dominated regions.

The spectra we have observed is characterized by having the exponent
close to $\alpha = 3 $. Thus, with reference to Sec. III this will
correspond to the spectrum in the anisotropic regime where the
stripes of vorticity gradients are near straight lines. Indeed in
Fig. 3 we see that we have stripes that are close to straight lines
and the observed spectrum is this in keeping with the expectations.
To illustrate the anisotropic nature of the spectrum directly we
plot in Fig. 6 the two dimensional spectrum, $\epsilon(k_x , k_y)$.
We observe a clear anisotropy, which become particular apparent in
the compensated spectrum in Fig. 6b, which is obtained by
subtracting the angle average of $\epsilon(k_x , k_y)$ (i.e.,
$(2\pi)^{-1}\int_{0}^{2\pi}\epsilon(k_x , k_y) d\theta $). Here we observe a
clear angular structure. We should emphasize that the spectra
obtained here are instantaneous spectra obtained a one time and for
one realization. Ensemble averaged spectra are expected to show a
much higher degree of --  if not complete --  angle isotropy.


\section{Concluding remarks}

We have performed a detailed investigation of the relation between
turbulent spectra and possible singularities in 2D turbulent flows.
First, we have demonstrated that the $k$-behavior of the spectra
generated by sharp vorticity gradients, based on the compressible
advection of di-vorticity, depends significantly on the anisotropy
of the spectra. If the angular spectrum distribution has one or more
very sharp peaks then the one-dimensional spectrum has a tail
falling-off like $k^{-3} $ at large $k$, which is resembling the
Kraichnan spectrum, derived from spectral cascade arguments. In the
opposite case of an isotropic smooth angular dependence the spectrum
has an  asymptotic behavior $k^{-3} $ as for the Saffman spectra.
These arguments allow us to suggest interpretation of many numerical
experiments where the  spectral exponent varies in the whole
interval between $-3$ and $-4$. For instance, in the simulations by
Okhitani \cite{okhitani} the {\it e}-regions, because of their
geometry, would give the main contributions to the isotropic
component of the spectrum which explain the Saffman exponents for
the observed spectrum in \cite{okhitani}. For {\it h}-regions the
situation is different: these regions contain stretched vorticity
level lines and their contribution to the spectrum should be
expected to be very anisotropic. This is why these regions produce
the $k$-behavior as for the Kraichnan spectrum. A similar situation
takes place in our simulations in comparison with the numerics
performed before \cite{NHRB,RNN}. In the both latter simulations the
spectra were isotropic resulting in spectral exponents like for the
Saffman spectrum. In the simulations presented in the present paper
we have very strong vorticity gradients concentrated on the very
narrow stripes and therefore the exponent is close to that for the
Kraichnan spectrum. Employing a filtering of the vorticity field
indicates, more or less one-to-one, that the tail of the spectrum
originates from the sharp vorticity gradients. A strong
amplification of di-vorticity of more than hundred times is one of
the main results of our simulations. A detailed investigation of the
growth of the di-vorticity maximum is beyond the main scope of this
paper and will be considered in future works. In conclusion, we
stress that this amplification has a natural explanation due to
compressibility of the mapping (\ref{map1}) providing the transfer
from the Eulerian description to the system of movable curvilinear
di-vorticity lines as described in Sec. II.

\section{Acknowledgments}

This work was supported by INTAS (grant no. 00-00292). The work of
E.K. was also supported by the RFBR (grant no. 00-01-00929). E.K.
wishes to thank Riso National Laboratory, where this work was
initiated, for its kind hospitality during the visit in November,
2003. JJR thanks the Landau Institute for kind hospitality during a
visit in June 2004, where the work was finalized.

\newpage

\section*{Figures}

\begin{figure}[h]
\centering a)
\includegraphics[width=10.0cm]{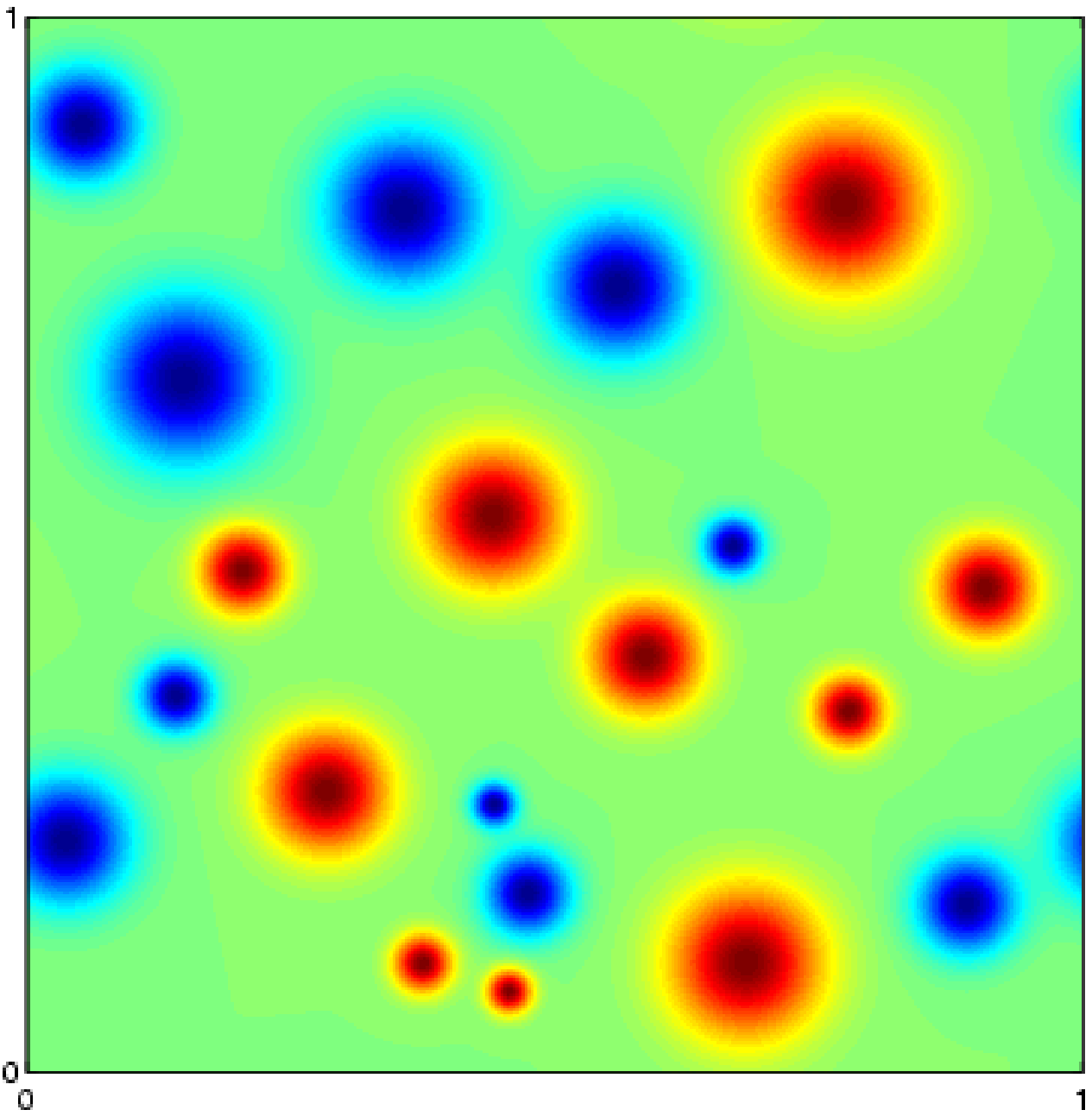}\\
\vspace{0.5cm} b)
\includegraphics[width=10cm]{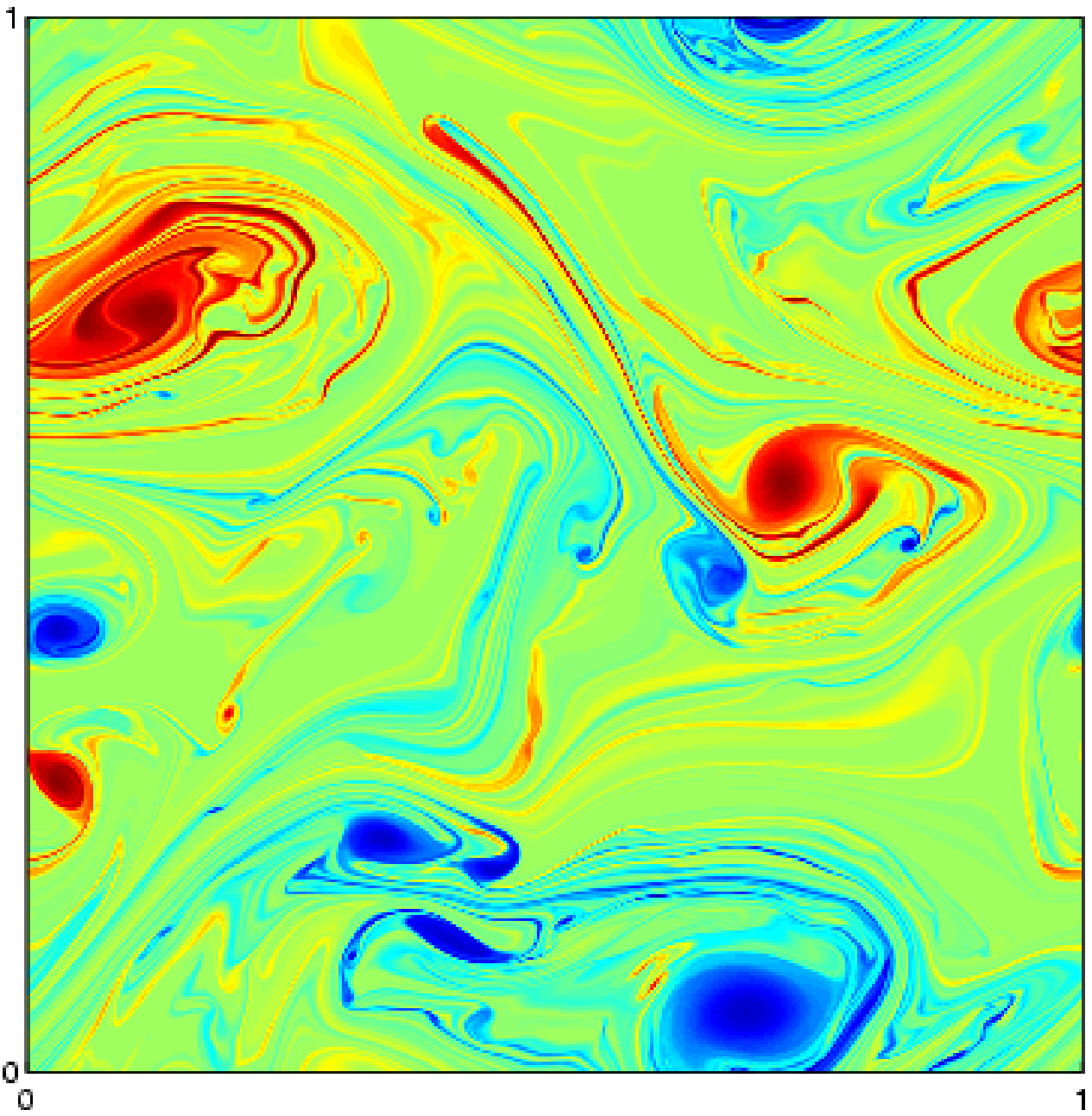}\\
\caption{ a) Initial vorticity field. b) Vorticity field at time 95
corresponding to 10 vortex turnover times. Red color designates
positive vorticity and blue color negative vorticity; maximum value
is 1 and minimum value is -1. $\omega_0 = 1$} \label{Fig:Vorticity}
\end{figure}

\vspace{1.0cm}

\begin{figure}[h]
\centering
\includegraphics[width=15cm]{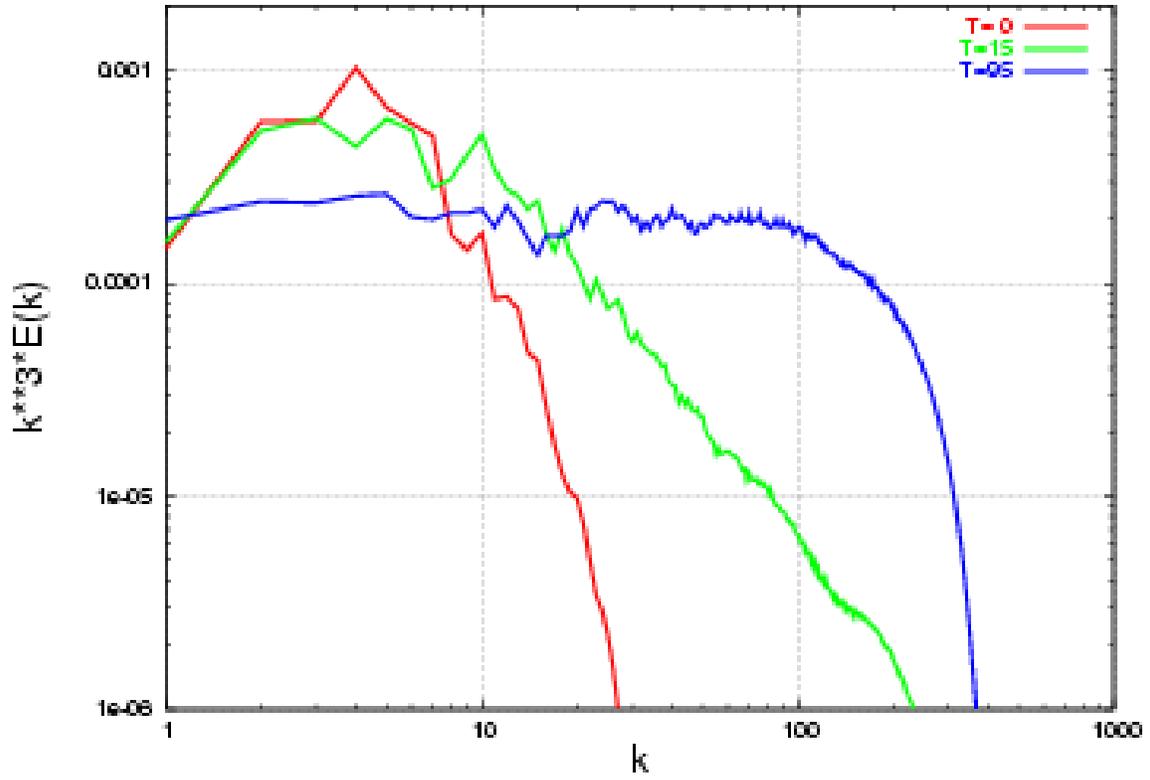}
\caption{ Compensated energy spectrum at different times $k^3
E(k)$ corresponding to the vorticity field shown in Fig.~\ref{Fig:Vorticity}.}
\label{Fig:spectra}
\end{figure}

\vspace{1.0cm}

\begin{figure}[h]
\centering
\includegraphics[width=10cm]{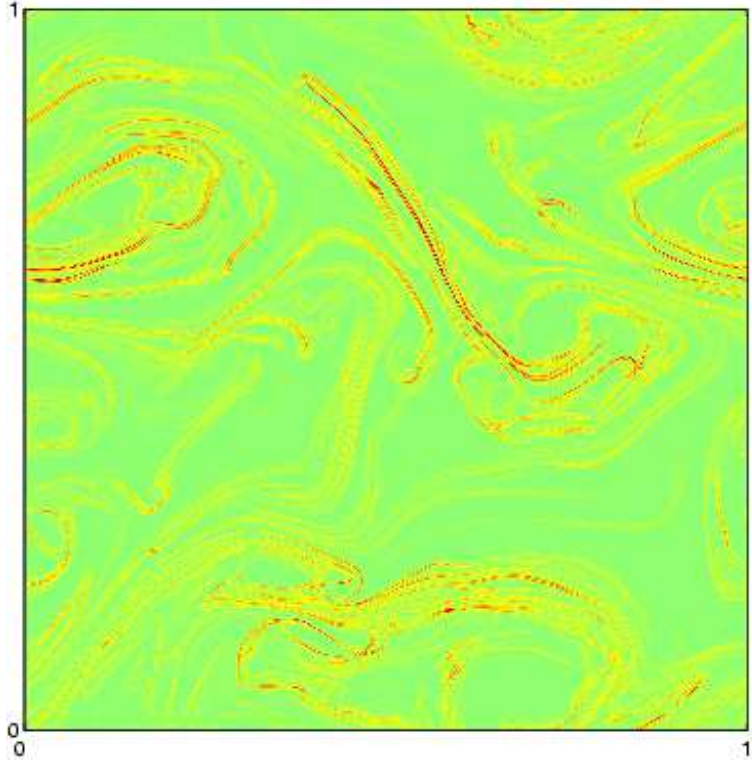}
\caption{The di-vorticity field ${\bf B}$ at $T = 95$. Here the length
of di-vorticity vector with the maximum (red) value being 673.} \label{Fig:divorticity}
\end{figure}

\vspace{1.0cm}

\begin{figure}[h]
\centering
\includegraphics[width=10cm]{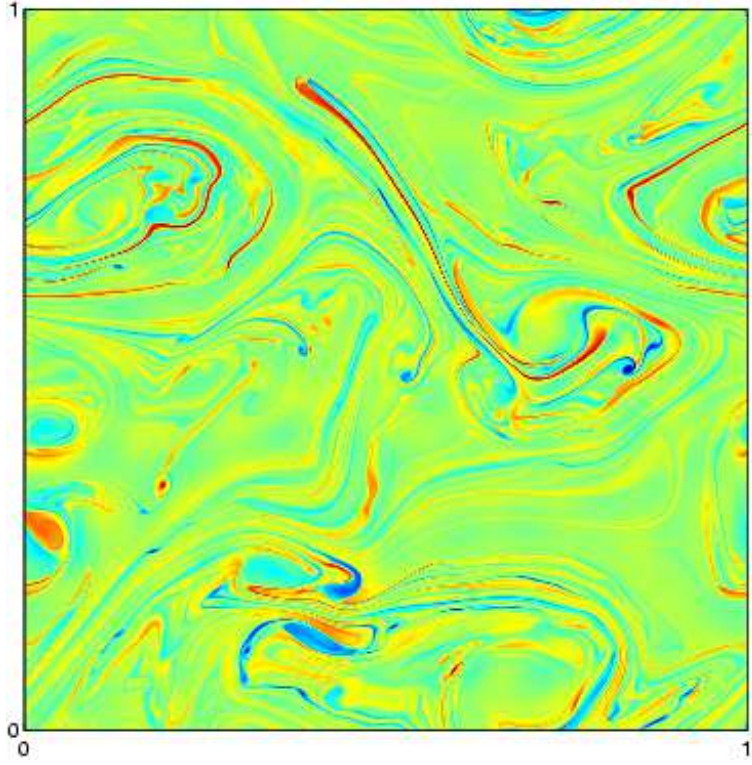}
\caption{ High pass filtered vorticity field from Fig. 1b, $k>10$.}
\label{Fig:Filt_vort}
\end{figure}

\vspace{1.0cm}

\begin{figure}[h]
\centering
\includegraphics[width=10cm]{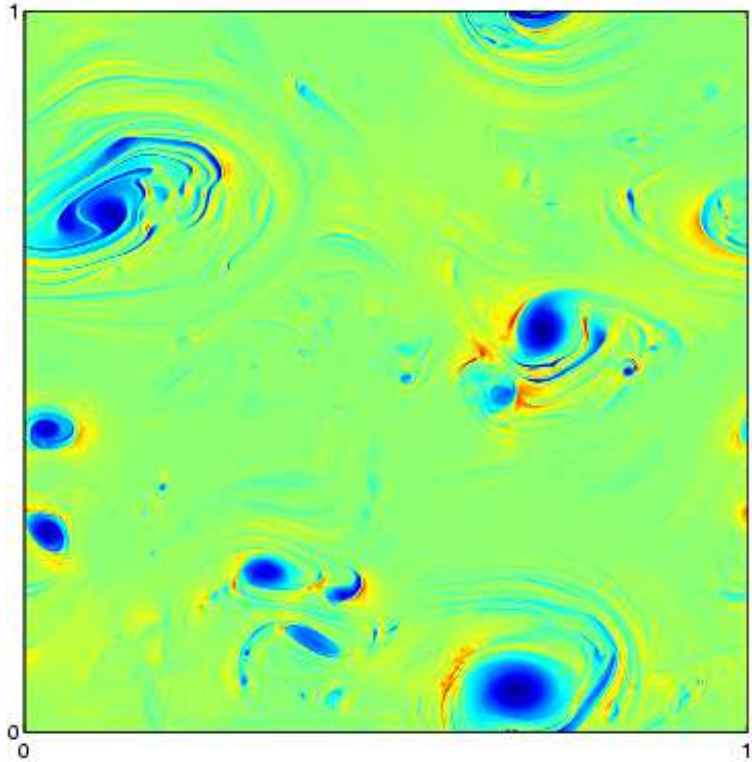}
\caption{  Weiss field for the vorticity field show in Fig.~1b. Red designates
positive values, i.e., strain dominated regimes. Blue designates
negative values, i.e., vorticity dominated regimes.}
\label{Fig:Weiss_Field}
\end{figure}

\vspace{1.0cm}

\begin{figure}[h]
\centering \centering a)
\includegraphics[width=14.0cm]{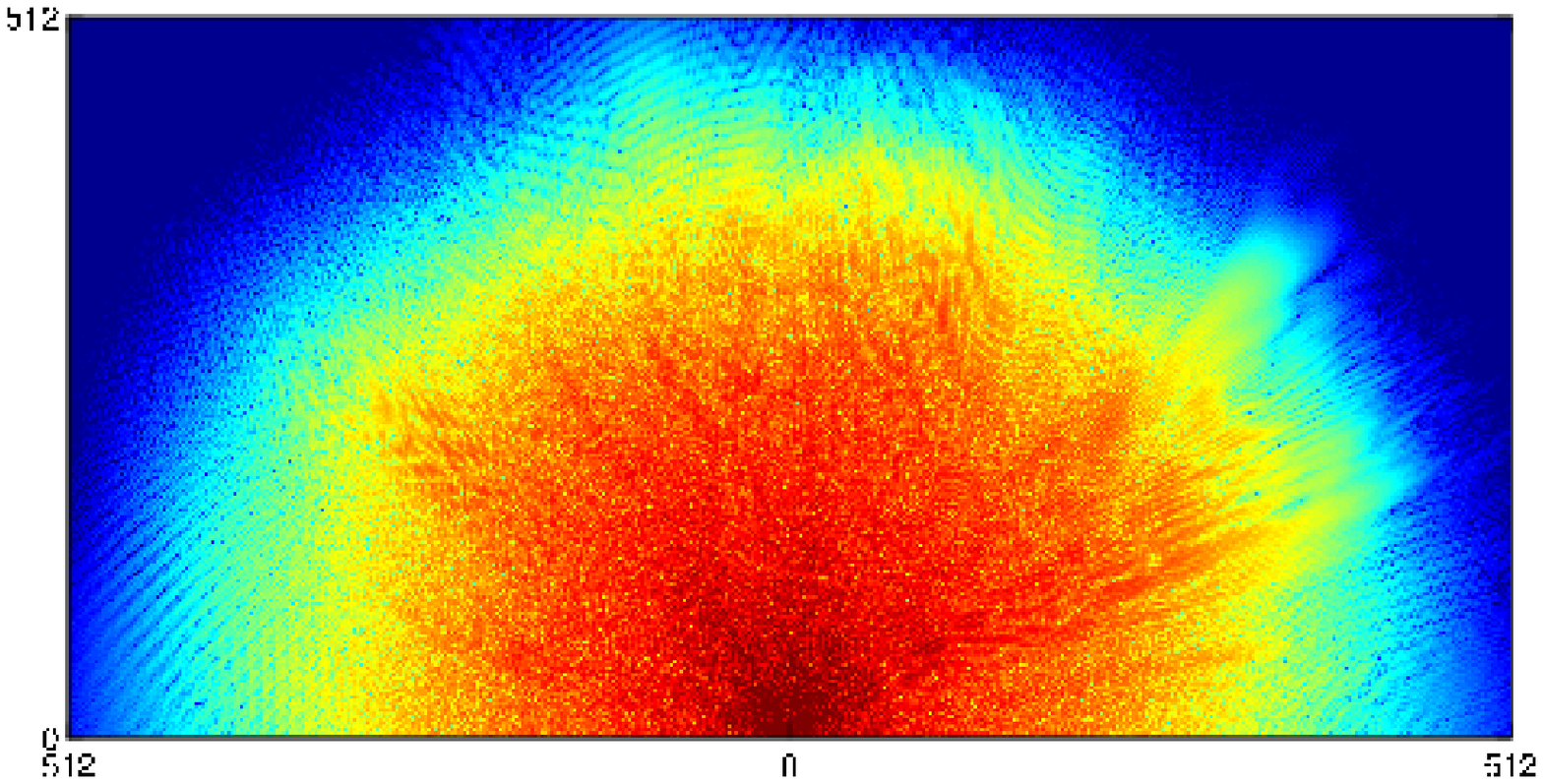}\\
\vspace{0.75cm} b)
\includegraphics[width=14cm]{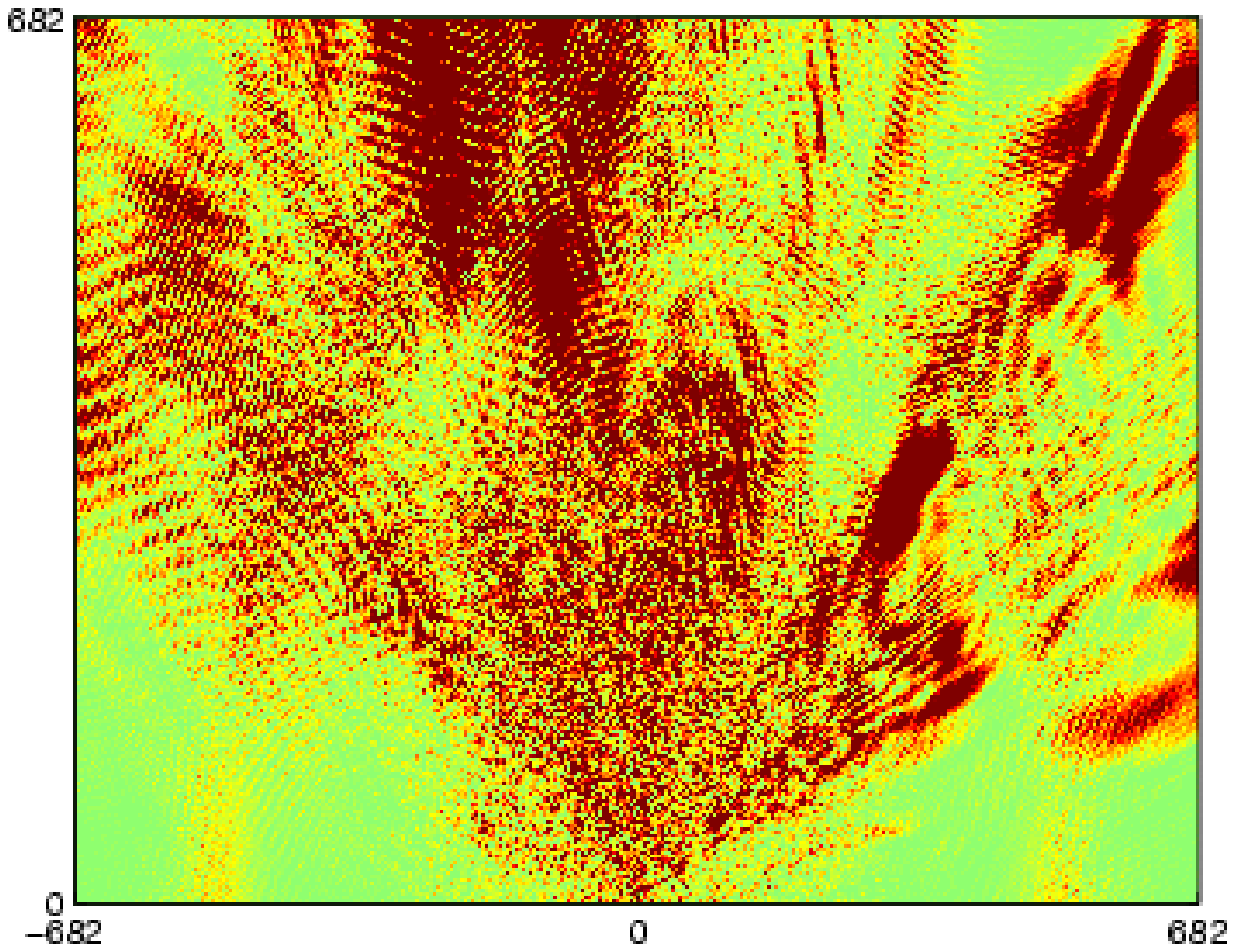}\\
\caption{a) Logarithm of the 2D energy spectrum $\ln \epsilon(k_x , k_y )$, logarithmic
Red designates the highest positive, blue the
negative values. b) The 2D energy spectrum
$k^4\epsilon(k_x , k_y )$ compensated by
subtracting the angle averaged spectrum, linear scale. Red
designates the highest values, blue is "zero". }
\label{Fig:2D_spek}
\end{figure}

\end{document}